\documentclass[a4paper, amsfonts, amssymb, amsmath, reprint, showkeys, twoside,superscript address]{revtex4-2}

\newcommand{\mez}{\hspace*{+0.50cm}}
\newcommand{\mz}{\hspace*{+0.25cm}}
\newcommand{\m}{\hspace*{-0.50mm}}
\newcommand{\n}{\hspace*{-0.25mm}}
\newcommand{\be}{\begin{equation}}
\newcommand{\ee}{\end{equation}}
\newcommand{\la}{\langle}
\newcommand{\ra}{\rangle}
\usepackage{bm,amssymb,amsmath}
\usepackage{graphicx,color,upgreek}
\usepackage{float}
\usepackage{xcolor}
\usepackage{relsize}
\usepackage{comment}

\begin{document}

\vspace*{+1.00cm}

 \title{Restoring heat and particle flow in strongly coupled non-equilibrium devices}

  \author{Noa Ludwin}
 \affiliation{Physics Department, Technion-Israel Institute of Technology, Haifa 3200003, Israel}
  \author{Ohad Cremerman}
 \affiliation{Physics Department, Technion-Israel Institute of Technology, Haifa 3200003, Israel}
\author{ Milan \v{S}indelka}
 \affiliation{ 
Department of Chemistry, Guangdong Technion Israel Institute of Technology,
Shantou, Guangdong Province 515603, People’s Republic of China}
\affiliation{ 
Schulich Faculty of Chemistry, Technion–Israel Institute of Technology, Haifa 32000, Israel}
 \author{David Gelbwaser-Klimovsky}
 \email{\tt dgelbi@technion.ac.il}
 \affiliation{Schulich Faculty of Chemistry and Helen Diller Quantum Center, Technion-Israel Institute of Technology, Haifa 3200003, Israel}

\begin{abstract}
Under non-equilibrium conditions, energy/particle currents flow through a quantum system coupled to multiple baths. Although the fluxes increase in the weak coupling regime as the coupling strengthens,  the reason why they decrease to zero for large coupling remains unknown. This counterintuitive behavior of energy exchange or transport phenomena is called the turnover effect. It has been predicted in several quantum systems—such as photosynthetic complexes, mesoscopic junctions, quantum heat machines, and chemical networks—without a single counterexample, and it results in constrained performance due to limited currents. Here, we use scattering theory to study the turnover effect produced by low-density reservoirs of free particles scattered by a quantum system through localized potentials. We find that the turnover effect is a consequence of total reflection that impedes the reservoir particle from reaching the interaction region and exchanging energy with the quantum system.  Moreover, we design a protocol based on quantum tunneling to avoid the turnover and to increase the maximum current. This strategy could be used to unleash the full potential of devices based on temperature/chemical potential gradients. 
\end{abstract}

\maketitle

\section{Introduction}

\begin{figure}[htbp]
		\centering
		\includegraphics[width=1\linewidth]{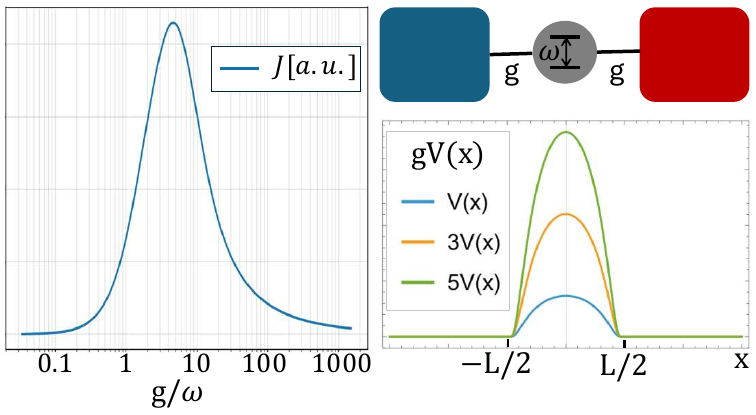}
		\caption{ Left: Turnover of the heat flow as a function of the normalized interaction strength, $g/\omega$. The general setup is depicted on the top right, where two thermal baths  or electrodes interact through an intermediate system. As an example, we consider a two-level system coupling two baths, each of them composed of a 1D free particle gas that is scattered by the system through a bump potential, $g V(x)=g e^{\frac{1}{x^2-(L/2)^2}}$ for $|x|<L/2$ and zero otherwise (bottom right). For  an example of the turnover in other realizations, see \cite{gelbwaser2015strongly,anto2021strong,wang2015nonequilibrium,batge2021nonequilibrium,segal2016vibrational,katz2016quantum,kato2015quantum,velizhanin2008heat,rips1990quantum,hanggi1990reaction,barsky2026impossibility,anto2023effective,tolkunov2007quantum,rebentrost2009environment,liu2019thermodynamic,kaneyasu2023quantum,segal2014two,liu2017energy}. See Supplementary Information for the simulation details and plot parameters.}
      \label{fig:1}	
	\end{figure}

In non-equilibrium scenarios, temperature or chemical potential gradients are used to generate heat or electric currents.  The current depends on the strength of the coupling between a mediating system and reservoirs (electrodes or thermal baths) at different chemical potentials/temperatures (see Fig. \ref{fig:1}).

Although in the weak coupling regime the energy/particle fluxes increase as a function of the system-reservoir coupling strength, the fluxes reach a maximum value at an intermediate strength, and decrease to zero for large coupling strengths \cite{gelbwaser2015strongly,anto2021strong,wang2015nonequilibrium,batge2021nonequilibrium,segal2016vibrational,katz2016quantum,kato2015quantum,velizhanin2008heat,rips1990quantum,hanggi1990reaction,barsky2026impossibility,anto2023effective}. Counterintuitively, for parameters where the system and the reservoirs strongly interact, they act as though they were completely decoupled from each other without exchanging energy or particles \cite{tolkunov2007quantum}.  We refer to this behavior as \emph{the turnover effect}. It has been predicted in a wide range of systems, such as photosynthetic complexes \cite{rebentrost2009environment}, thermoelectric devices \cite{liu2019thermodynamic}, superradiant systems \cite{tolkunov2007quantum}, and quantum heat machines \cite{gelbwaser2015strongly,kaneyasu2023quantum}. Moreover, chemical reactions also exhibit a turnover effect as a function of the interaction strength with the reservoirs, known as Kramers turnover \cite{rips1990quantum}. As a consequence of the turnover effect, the performance of several non-equilibrium setups is limited by a bound on  electrical current, heat flow, power, or reaction rate.  The effect seems not to be restricted to a specific type of physical realization. It has been predicted in a diverse group of physical systems and reservoirs, including fermionic \cite{segal2014two} and bosonic reservoirs \cite{wang2015nonequilibrium}, and for mediating systems such as two-level systems \cite{liu2017energy}, harmonic oscillators \cite{velizhanin2008heat}, and molecules \cite{katz2016quantum}.   To the best of our knowledge, there is no single example in the literature of a system strongly coupled to multiple reservoirs that does not present the turnover effect.  The prevalence of the turnover effect across such a large and diverse group of physical setups, together with the lack of counterexamples, could be taken to indicate its universality. Furthermore, its ubiquity suggests a fundamental physical principle that has yet to be established  that applies across diverse systems. 

In this letter, we study the turnover effect using the scattering framework, in which scattering particles serve as low-density reservoirs. The interaction between the reservoirs is mediated by a quantum system that interacts through localized potentials that create an interaction region. We show that the turnover effect is a consequence of total reflection, which, in the strong-coupling limit, inhibits the access of scattering particles to the interaction region, thereby halting the current between reservoirs.  Furthermore, we develop a protocol that avoids the turnover and even increases the maximum current. This protocol could be used to create quantum switches and to enable the transfer of heat or electrical current in the strong-coupling regime.

\section{Turnover effect}
We start by analyzing the interaction of a quantum system with a single reservoir and its dependence on the system-bath coupling strength. This analysis is simpler than the multi-bath setup and, as shown below, provides the basic intuition needed to explain the turnover effect.  In particular, we will consider the bath to be  a gas of free particles of mass $m$  which are scattered by an $N-$level system. 
$\hat{H}_S=\sum_{j=1}^N \varepsilon_j|j\rangle\langle j|$ is the free Hamiltonian of the $N-$level system and it is bounded.  The interaction Hamiltonian is given by $\hat{H}_{int}=g\sum_{i=1}^N V_i(\mathbf{r})|\chi_i\rangle \langle\chi_i| $, where $\mathbf{r}$ is the position of a particle and $\{|\chi_i\rangle\}$ is an orthonormal eigenbasis in the $N-$level system Hilbert space. Notice that if $|\chi_i\rangle$ are not eigenfunctions of $\hat{H}_S$, energy could be transferred between a particle and the $N-$level system. $g$ controls the strength of the interaction. Standard scattering assumptions require the particle and the $N-$level system to be completely independent for large separations. We achieve this by assuming that outside a finite interaction region,  $\Omega \subset {\mathbb R}^D$, the potentials $V_i(\mathbf{r})$ are zero.  Even more specifically,
 we assume that the potentials are repulsive inside the interaction region, that is,  $V_i(\mathbf{r})>0$ for $\mathbf{r}\in \Omega$.
We start by studying the scattering of a single particle by the quantum system. In this case, the  corresponding time dependent Schr\"odinger equation   reads as $i \hbar \partial_t |\Psi (t)\ra=\hat{H}|\Psi (t)\ra$. Here $|\Psi (t)\ra$ is a square integrable wavepacket, and
\begin{equation}
\hat{H}=  -\frac{\hbar^2}{2m}\nabla^2+g\sum_{i=1}^N V_i(\mathbf{r})|\chi_i\rangle \langle\chi_i|+\hat{H}_S
\;; \label{eq:scheq}
\end{equation}
 with $\nabla$ operating on the particle's position. 
The wavepacket can be expanded as $|\Psi(t)\rangle=\sum_{i=1}^N\psi_i(t,\mathbf{r})|\chi_i\rangle$. Calculating the  expectation value of $\hat{H}$  over the state $|\Psi(t)\rangle$, 
we get
    	\begin{figure}[htbp]
		\centering
\includegraphics[width=1\linewidth]{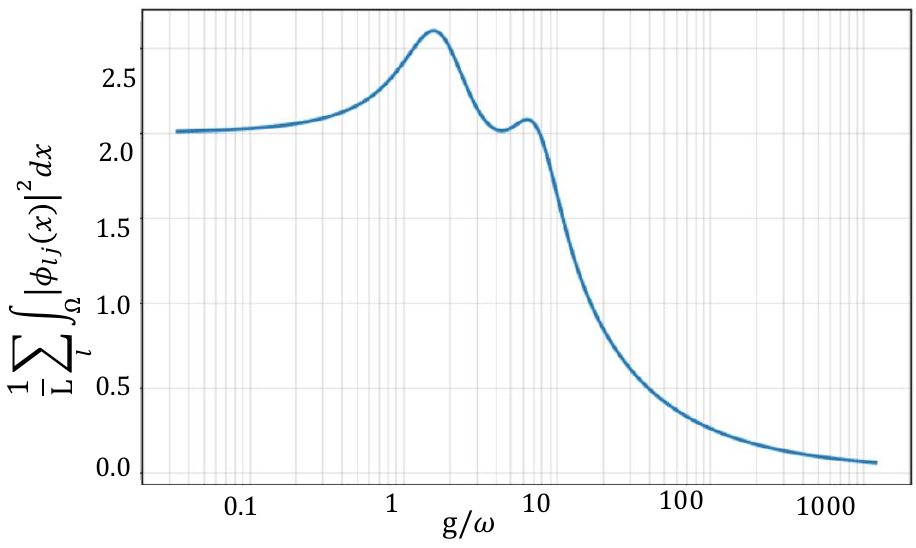}
		\caption{  Squared amplitude of the scattered wavepacket integrated over 
        the interaction region and plotted as a function of the normalized coupling strength. We consider the same realization as in Figure 1. The wavepacket  takes the form $|\Psi\rangle=\sum_{l=1}^{N}\phi_{l j}(x)|l\rangle$, where $|l\rangle$ are  the $\hat{H}_S$ eigenstates, and the asymptotic  in-state of the system is $|j\rangle$. The mean wavepacket energy $E$ is specified in the Supplementary Information. For wavepackets with other energies, the amplitude in the interaction region also tends to 0 as $g$ grows (see the Supplementary Information for all the plot parameters).} \label{fig:2}
	\end{figure}

\begin{gather}
E=  \langle \Psi(t)| \hat{H}|\Psi(t)\rangle
=  \langle \Psi(t) |\left(-\frac{\hbar^2}{2m}\nabla^2+\hat{H}_S\right)|\Psi(t)\rangle\notag \\
+g\sum_i
\int_\Omega d^D\hspace*{-0.50mm} r \, |\psi_i(t,\mathbf{r})|^2 \, V_i(\mathbf{r}) \;.\label{eq:scheqpro1}
\end{gather}
Notice that the first term on the r.h.s.~is the expected value of the kinetic energy, which for any physical system has to be finite and positive. The same holds for $E$. 
The second term is also finite because the spectrum of $H_S$ is bounded. The only remaining  contribution is the interaction term
\begin{gather}
    g\sum_i
\int_\Omega d^D\hspace*{-0.50mm} r \, |\psi_i(t,\mathbf{r})|^2 \, V_i(\mathbf{r}) \; .
\label{eq:div}
\end{gather}
Here we use the fact that the potential is zero outside the interaction region $\Omega$. Due to our restriction to repulsive potentials, this term is positive. It might be thought that, as $g\rightarrow \infty$, this term diverges. But  such behavior would be inconsistent with  Eq. \eqref{eq:scheqpro1}, since none of the other terms diverge.  
Such a
 contradiction can be prevented only if the wavefunction inside the interaction area is zero, i.e., $|\psi_i(t,\mathbf{r})|^2=0$ for $\mathbf{r}\in \Omega$ (see Figure \ref{fig:2}).   Physically, this implies that for $g\rightarrow \infty$ the particle can not enter the interaction region and therefore will not be able to exchange energy with the quantum system. Whenever the particle reaches the interaction area, it is perfectly reflected.

 Notice that, for attractive potentials, $V_i(\mathbf{r})<0$,  the quantity \eqref{eq:div} will be negative. Deep attractive potentials may produce large kinetic energy that may compensate a large negative term in Eq. \eqref{eq:div}. Here, we excluded these cases by restricting ourselves to repulsive potentials. We study the attractive case  below using an analytic toy model, which suggests that, for these potentials as well, the scattered particle does not enter the interaction region as $g\rightarrow\infty$ (see Section III.A). 

Now that we understand the scattering of a single particle  in the limit of $g\rightarrow\infty$, we can study the case of a reservoir of particles and also of multiple reservoirs that could produce non-equilibrium conditions such as gradients of temperatures or chemical potentials.  If the particle density is low, any scattering event is a single-particle  process. Therefore,
 in the low density regime, our analysis above still holds, and there will be no energy transfer at $g\rightarrow\infty$ for a single or multiple reservoirs.  This explains the turnover effect. 

The analysis  presented above also allows us to design a strategy to avoid the turnover effect. As far as we are aware, there is no single example in the literature of energy or particle flow in a non-equilibrium system where the current does not decrease with the coupling strength in the limit of $g\rightarrow\infty$. 

\section{Avoiding the turnover effect}

Avoiding the turnover effect could be especially relevant for non-equilibrium devices that must operate in the strong-coupling regime.  Moreover, as we show below,  the proposed protocol not only allows the exchange of heat/particles at the strong coupling, but it could increase the maximum heat/particles flow. 

As explained above, the turnover originates due to necessity to avoid the divergence in Eq. \eqref{eq:scheqpro1}. But  divergence in
Eq. \eqref{eq:scheqpro1} can be avoided  also by having a zero interaction, $V_i(\mathbf{r})=0$  for some $i$, instead of a zero wavefunction.  Therefore,  a state $|\chi_i\rangle$ which is not part of the interaction Hamiltonian will allow a non-zero $\psi_i(\mathbf{r})$ for $ \mathbf{r}\in\Omega$. This implies that the scattering particle could enter the interaction region and exchange energy with the quantum system. 

This situation can be achieved as follows.  We start by adding an energy level to the $N-$level system.  It is essential not to change the interaction Hamiltonian by adding this level. 
Below, we provide a potential physical implementation. Adding an energy level increases the Hilbert space and, therefore, the size of the eigenbasis.  Considering  a wavepacket in the new basis, we get $|\Psi (t)\rangle=\sum_{i=1}^{N+1}\psi_i(t,\mathbf{r})|\chi_i\rangle$.  We have already established that in the interaction region $\psi_i(t,\mathbf{r})=0$ for $i\leq N$, but  because $V_{N+1}(\mathbf{r})=0$, there is no constraint on $\psi_{N+1}(t,\mathbf{r})$. The pertinent expectation value of $\hat{H}$ over the state $|\Psi (t)\rangle$ is
\begin{gather}
E=  \langle \Psi(t)| \hat{H}|\Psi(t)\rangle
=  \langle \Psi(t)| \left(-\frac{\hbar^2}{2m}\nabla^2+\hat{H}_S\right)|\Psi(t)\rangle\notag \, .\label{eq:scheqpro2}
\end{gather}
Here, the potential divergence problem of Eq. \eqref{eq:scheqpro1} disappears, so $\psi_{N+1}(t,\mathbf{r})$ can be different from zero also in the interaction region, allowing the gas particle to interact with the quantum system, exchange energy with it, and potentially avoid in this manner the turnover effect.  

\subsection{Analytical model}

To test whether the turnover persists after adding an extra energy level, we study a toy model that can be solved analytically.  For simplicity, we  consider 
 gas particles moving only in 1D and  take
 the interaction potentials  to be proportional to Dirac delta functions, such that $\hat{H}_{int}=g\sum_{i=1}^{N} v_i\delta(x-x_i)|x_i\rangle \langle x_i| $. Here, $v_i$ can take any value, so this model also works for attractive potentials.
Vectors $|x_i\rangle$ represent states localized at points $x_i$.  As an example of physical realization, consider a single electron that can tunnel among $N$ sites (e.g., quantum dots), each of them localized at a different $x_i$ \cite{delgado_theory_2007}. At each site, only a single  electronic level is allowed. Due to tunneling $[\hat{H}_S,\hat{H}_{int}]\neq\hat{0}$, energy transfer is  enabled, and, as we show below, the turnover effect may be avoided. 
Notice that here the interaction region is just the set of points $\{x_i\}$, which are contained in $\Omega$. Furthermore, we assume that the separation between the sites is smaller than the de Broglie wavelength of the scattered particle, so we
can consider that all the interaction sites being centered at the origin, so $\hat{H}_{int}\approx g\delta(x)\sum_{i=1}^Nv_i|x_i\rangle\langle x_i|\equiv g\delta(x)\hat{S}_N$.  This allows us to obtain analytical expressions for $|\Psi\rangle $ and study its behavior as a function of the interaction strength $g$.

To find $|\Psi\rangle$  we use the Lippmann-Schwinger equation (LSE) \cite{LSE,LSE-2,LSE-3}: $|\Psi\rangle=|pj\rangle+\hat{G}_0\hat{H}_{int} |\Psi\rangle$,  where $\hat{G}_0$ is the free retarded Green's operator and $|pj\rangle$ is the  pertinent scattering in-state, which is an eigenstate of the free Hamiltonian, $\hat{H}_0=\frac{\hat{p}^2}{2m}+\hat{H}_S$. Due to the form of the LSE, it is worth expanding $|\Psi\rangle$ in the eigenbasis of $\hat{H}_S$ as $|\Psi\rangle= \sum_{l=1}^{N}\phi_{l j}(x)|l\rangle$. Notice the double index on $\phi_{lj}(x)$, where the first index corresponds to the basis expansion and the second to the incoming scattering channel. Because we are interested in transition rates between different states we focus here on $l\neq j$. After solving the LSE we  find that $\phi_{lj}(x)$ at the interaction point, $x=0$, is a ratio of polynomials in $g$ (see  Supplementary Information):
\begin{equation}
    \phi_{lj}(0)=\frac{\sum_{i=1}^{N-1} n_i g^i}{\sum_{i=0}^{N-1} u_i g^i+C Det[\hat{S}_N]g^N} \; .\label{eq:wavef}
\end{equation}
Here $n_i,$  $u_i$ and $C$  are functions of $E=\frac{p^2}{2m}+\varepsilon_j$, the incoming scattering channel energy. 
If all the sites interact with the particle gas, $v_i\neq 0$, then $Det[\hat{S}_N]\neq0$, and for strong coupling the wavefunction in the interaction region decays at least as $1/g$. From  Eq.~(\ref{eq:wavef}), we can also analyze the protocol that we previously proposed to avoid the turnover.  Namely, assume that we start with an $N-$level system with $\hat{H}_{int}=g\delta(x)\hat{S}_{N}$. Adding a level without changing $\hat{H}_{int}$ is equivalent to having $\hat{S}_{N+1}=v_{N+1}|x_{N+1}\rangle\langle x_{N+1}|+\hat{S}_{N}$ with $v_{N+1}=0$. In this case, $Det[\hat{S}_{N+1}]=0$, so the polynomial in the denominator is at most of order $N$, and the polynomial in the numerator becomes of the same order. Therefore, $\phi_{lj}(0)$ may get a finite non-zero value as $g\rightarrow\infty$ (see Supplementary Information), potentially allowing  heat currents in this limit. Moreover, as a consistency check, it can be shown that  $\phi_{lj}(0)=0$ for $g=0$.

To confirm that our protocol actually avoids the turnover effect, we calculate the heat flow in a non-equilibrium setup. For this, we consider the following  concrete toy model:  a non-degenerate two-level system with Bohr frequency $\omega$ interacting with two different gases at different temperatures. The interaction with the hot particles is given by $\hat{H}^h_{int}=g\sum_{i=1}^{2} v^h_i\delta(x-x_i)|x_i\rangle \langle x_i|$ and with the cold particles by $\hat{H}^c_{int}=g\sum_{i=1}^{2} v^c_i\delta(x-x_i)|x_i\rangle \langle x_i|$. For simplicity, we assume $v_i^h=v_i^c=v_i$.

    	\begin{figure}[htbp]
		\centering
		\includegraphics[width=1\linewidth]{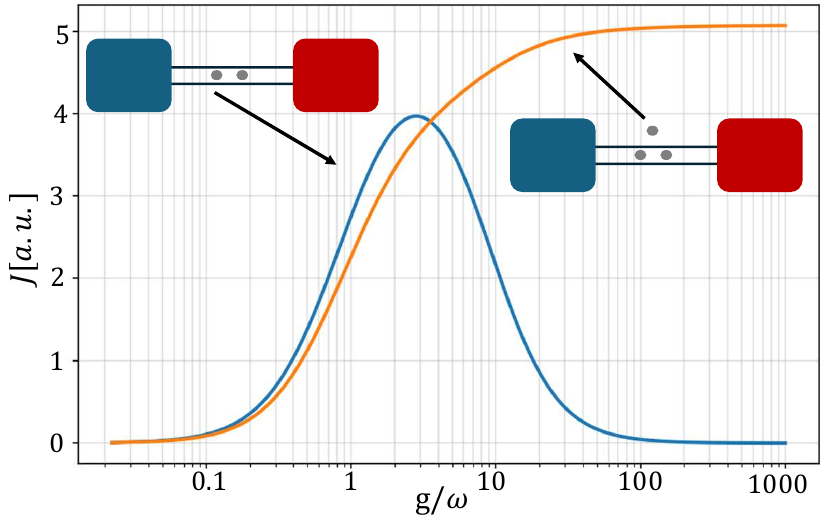}
		\caption{Heat flow as a function of the normalized coupling strength for a two-site intermediate system (blue line) vs.~three-site intermediate system (yellow line). The sites are represented as gray circles and the system-reservoir interaction region is delimited by black lines. All the sites  of the two-site system  are in the interaction region, whereas in the three-sites system, one is outside;  an electron on the outside site does not interact with the reservoirs but can still tunnel to the other sites. The presence of the third site outside the interaction region prevents the turnover (see the Supplementary Information for the plot parameters). }	\label{fig:3}
	\end{figure}
 The  setup can be realized using two quantum dots or sites (see left sketch in Figure \ref{fig:3}). We assume that the  gas particle number density is low, so the particles do not interact with each other, and the reduced dynamics is given by the low-density limit Lindblad equation \cite{dumcke_low_1985,alicki_violation_2023}.  Therefore, populations and coherences decouple and the latter decay. The dynamics of the populations, $p_l$, is governed by a Pauli rate equation, $\dot{p}_l=\sum_{j\neq l} \left(a_{lj}p_j-a_{jl}p_l\right)$. Here, $a_{lj}$ is the total transition rate from state $j$ to $l$. $a_{lj}$  is a linear combination of the rates produced by each bath, i.e., $a_{lj}=a^c_{lj}+a^h_{lj}$, where (see Supplementary Information)
\begin{equation}
    a^k_{lj}=\frac{2\hbar \nu_k}{Z_k}\int_{Max \{\varepsilon_l,\varepsilon_j\}}^\infty dE e^{-\beta_k\left(E-\varepsilon_j\right)}\sqrt{\frac{E-\varepsilon_l}{E-\varepsilon_j}}
    \left|\phi^k_{lj}(0)\right|^2 \; .\label{eq:rates}
\end{equation}
Here $k \in \{c,h\}$ labels the bath, $\nu_k$, $\beta_k$ and $Z_k$ are its density, inverse temperature and partition function, respectively.  $\phi_{lj}^k(0)$ is calculated considering only the interaction with the $k$-th bath. Eq. \eqref{eq:rates} shows that if the wavefunction is zero in the interaction area, the transition rates are also zero and no energy is exchanged between the bath and the quantum system.

The fact that the reduced dynamics is described by the  Lindblad equation allows us to write a thermodynamically consistent heat current as a function of the rates \cite{gelbwaser-klimovsky_thermodynamics_2015}. Although here we calculate heat currents produced by a temperature gradient, a similar calculation can be done for chemical potential gradients that will produce particle/electric currents. Therefore, our results on turnover also apply to electric currents.  The steady state heat current from bath $k$ takes the form
\begin{equation}
    J^k=\Sigma_{l>j}\left(\varepsilon_l-\varepsilon_j\right)\left(a_{lj}^kp_j^{ss}-a_{jl}^kp_l^{ss}\right),
\end{equation}
where $p_l^{ss}$ is the steady state population of level $l$. This can be found by setting $\dot{p}_l^{ss}=0$ on the Pauli rate equation with the total transition rates, $a_{lj}$. Because $Det[\hat{S}_2]\neq0$, the heat current exhibits a turnover (see blue line in Figure 3). We term $Max[J^h_2(g)]$  as the heat current
maximized over $g$. This is reached at $g/\omega\sim1$. 

Next, we add a third level without changing the interaction Hamiltonian. This can be achieved by adding a third site outside the region accessible to the particle gases (see right sketch in Figure 3). The site must be close enough to the other sites so there could still be tunneling to/from them. Even though the third site does not interact with any of the gases, tunneling  allows to effectively turn off the turnover by expanding the Hilbert space and getting $Det[\hat{S}_3]=0$. Due to the lack of turnover, the  maximized heat current, $Max[J^h_3(g)]$, is reached  at $g\rightarrow\infty$. In addition to avoiding turnover, the maximum heat flow for the three-level case can be larger than for the two-level case (see Figure 4). Notice that this is not a consequence of having larger Bohr frequencies. We avoid this by choosing the third level energy between the original two levels. The enhancement of the maximum current can also be achieved in an $N-$level system. Namely, consider an $N-$level system and add an extra level without changing the interaction as we did in the case of the two-level system, i.e., assume $v_{N+1}=0$.  By not changing $\hat{H}_{int}$  we ensure to get $Det[\hat{S}_{N+1}]=0$. Moreover, if we consider cases where  for $i\leq N$  the $v_i$ have approximately the same value, $v_i\approx c$, then $\hat{S}_{N}\approx c I$ and therefore, for the $N$-level system, $[\hat{H}_{int},\hat{H}_S]=\hat{0}$ for any $g$, preveting any energy exchange or current flux. The addition of an extra level  without changing $\hat{H}_{int}$  breaks the commutation relation and will  thus allow  energy exchange and heat transfer in non-equilibrium settings. 
    	\begin{figure}[htbp]
		\centering		\includegraphics[width=1\linewidth]{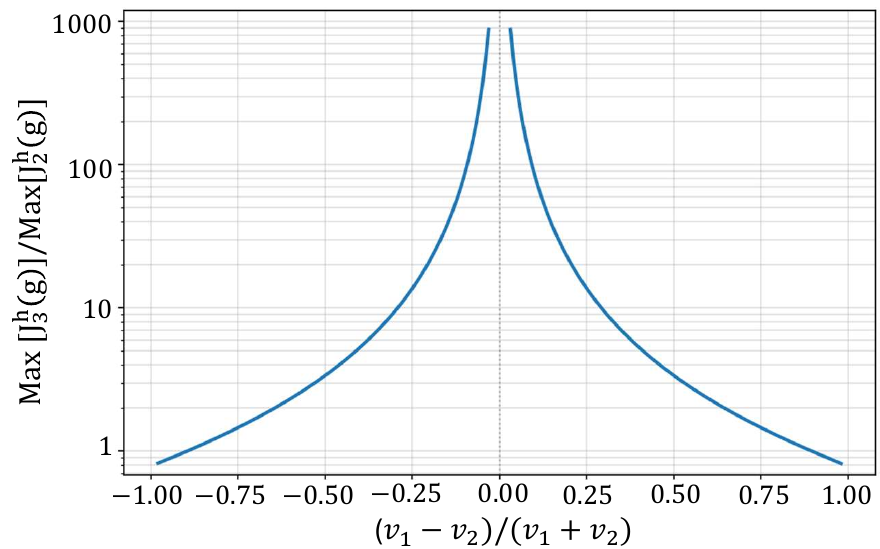}
		\caption{Ratio between maximum currents for the two and three sites described in Figure 3. For both cases, the  maximization of current is done with respect to the coupling strength
         $g$. A similar effect can be observed by increasing the number of levels from $N$ to $N+1$.}	
	\end{figure}

 Summarizing, we  have shown that, in scattering realizations, the turnover effect arises from total reflection, which prevents the scattered particle from entering the interaction region. Future work should clarify whether effects related to total reflection could help to explain the turnover effect in other realizations (e.g., spin-boson), where spatial localization of interactions is not explicitly indicated but still plays a role. An example of this hidden role of position can be found in collisional or repeated-interaction models, which, for simplicity, originally ignored particle positions but were improved once particle positions were accounted for \cite{gaida2025thermodynamically,tabanera2022quantum}. Finally, we showed that the turnover could be avoided if the interaction Hamiltonian has at least one zero eigenvalue. In these cases, not only is the turnover avoided, but the presence of zero eigenvalue can also increase the maximum current, enabling the creation of a tunneling-based quantum switch. An eventual generalization of this insight to other setups, such as spin-boson systems, will be the subject of future theoretical work.  



%

\onecolumngrid

\setcounter{equation}{0}
\renewcommand{\theequation}{S\arabic{equation}}

\setcounter{figure}{0}
\renewcommand{\thefigure}{S\arabic{figure}}

\setcounter{section}{0}
\renewcommand{\thesection}{S\arabic{section}}

\setcounter{subsection}{0}
\renewcommand{\thesubsection}{\thesection.\arabic{subsection}}
\vspace{1cm}

\begin{centering}
{\large \bf Supporting Information}\\
\end{centering}

\section{Dependence on $g$ of the wavefunction for delta potentials model}
In this section, we derive the dependence on $g$ of the wavefunction inside the interaction region for the case of a 1D delta potential (see equation 4 in the main text). We start with a derivation of the wavefunction and then study its dependence on $g$.

The Hamiltonian $\hat{H}_0=\hat{H}_S+\hat{H}_P$ of our two isolated subsystems satisfies
      \be
         \hat{H}_0 \, | p \,\n j \ra \; = \; \Bigl( E_p + \varepsilon_j \Bigr) \, |p \,\n j \ra ;
      \ee
      where $E_p=\frac{p^2}{2\,m}$. The interaction between the two subsystems is 
      
      \be    \hat{H}_{int}=g\delta(x)\sum_{i=1}^Nv_i|\chi_i\rangle\langle \chi_i|=g\delta(x) \hat{S}_N
      \ee
      It can be rewritten as 
      \be \label{coupling}
         \la p' \,\n l |\hat{H}_{int} | \Psi \ra \; = \;  \, \sum_{k}  \, v_{lk} \, \phi_{kj}\n(0)
         \; \frac{1}{(2\,\pi\,\hbar)^{1/2}};
      \ee

where

\begin{equation}
     v_{lk} =g \sum_{i}\la l|\chi_i\ra v_i \la \chi_i|k\ra.
\end{equation}

The implicit Lippmann-Schwinger equation (LSE)

         \be \label{LSE-implicit}
         | \Psi \ra \; = \; | p \,\n j \ra \; + \; G_0 \, \hat{H}_{int} \, | \Psi \ra.
      \ee
      Here $G_0$ is the free Green's operator, $G_0=\frac{1}{E_p + \varepsilon_j -\hat{H}_0 + i\epsilon}$. 

      Moreover, we can expand $|\Psi\rangle= \sum_{l=1}^{N}\phi_{l j}(x)|x l\rangle$ where $ \phi_{l j}(x) \; = \; \la x \,\n l | \Psi \ra  $. In a similar way, we can define $\mez \tilde{\phi}_{lj}(p') \; = \; \la p' \,\n l | \Psi \ra $. Using the LSE, we get

        \begin{eqnarray} \label{LSE-implicit-take-2}
      \tilde{\phi}_{lj}(p') & = & \delta\n( p-p'\n) \; \delta_{jl} \; + \;
         \int \m d\n p'' \sum_{k} \, \frac{\delta\n( p'\n- p''\n) \;
         \delta_{l\n k\n}}{E_p + \varepsilon_j - E_{p'\n} - \varepsilon_{l\n} + i\epsilon} \;
         \la  p''\n \,\n k\n | \hat{H}_{int} | \Psi \ra \; = \nonumber\\
         & = & \delta\n(p- p'\n) \; \delta_{jl} \; + \; \frac{\la  p'\n \,\n l\n | \hat{H}_{int} | \Psi \ra}
         {E_p + \varepsilon_j - E_{p'\n} - \varepsilon_{l\n} + i\epsilon} \; = \nonumber\\
         & = & \delta\n(p-p'\n) \; \delta_{jl} \; + \; \frac{1}{E_p + \varepsilon_j - E_{p'\n} - \varepsilon_{l\n} + i\epsilon}
         \; \sum_{k}  \, v_{l\n k} \, \phi_{kj}\n(0) \; \frac{1}
         {(2\,\pi\,\hbar)^{1/2}}.
      \end{eqnarray}
      where we used that

      \be \label{G-0-momentum-representation}
         \la p'\n \,\n l\n | \, \frac{1}{E_p + \varepsilon_j - \hat{H}_0 + i\epsilon} \, | p''\n \,\n k\n \ra \; = \;
         \frac{\delta\n(p'\n-p''\n) \; \delta_{l\n k\n}}{E_p + \varepsilon_j - E_{p'\n} - \varepsilon_{l\n} + i\epsilon}.
      \ee
Using 

      \be
           \phi_{lj}(0)  = \; \int \m \n d p' \;    \tilde{\phi}_{lj}(p') \;
         \frac{1}{(2\,\pi\,\hbar)^{1/2}};
      \ee
      and (\ref{LSE-implicit-take-2}) provides immediately
      \begin{equation} \label{eq:linsys}
        \phi_{lj}(0)  =  \frac{1}{(2\,\pi\,\hbar)^{1/2}} \; \delta_{jl}
           -\,\frac{i}{2\hbar}\,\sqrt{\frac{2\,m}{E-\varepsilon_{l\n}\n}} \sum_{k}  \,  v_{l\n k}   \phi_{kj}\n(0).
      \end{equation}
where we used that 

      	\be \label{integral-1D}
	\int_{-\infty}^{+\infty} \m {\rm d}p'\n \; \frac{1}{E_p + \varepsilon_j - E_{p'\n} - \varepsilon_{l\n} + i\epsilon} \; = \;
	-\,i\,\pi\,\sqrt{\frac{2\,m}{E-\varepsilon_{l\n}\n}};
	\ee
      
Rearranging \eqref{eq:linsys}, we get a set of coupled non-homogeneous linear equations.

\begin{equation}
    \sum_{k}\left(\delta_{lk}+\frac{i}{2\hbar}\,\sqrt{\frac{2\,m}{E-\varepsilon_{l\n}\n}}  \,  v_{l\n k} \right)  \phi_{kj}\n(0)  =  \frac{1}{(2\,\pi\,\hbar)^{1/2}} \; \delta_{jl}    
\end{equation}

The solution of the above system is the particle wavefunction at the interaction region. We are interested in the transition rates $a_{ij}$ for $i \neq j$. As we explain below, $a_{ij}$ depends on $ \phi_{ij}(0)$.  We can solve the system using Cramer's law. For $i \neq j$ 

\begin{equation}
 \phi_{ij}(0)=\frac{Det[A_i]}{Det[B]}   
\end{equation}
where $B$ and $A_i$ are matrices with elements
$(B)_{lk}=\delta_{lk}+\frac{i}{2\hbar}\,\sqrt{\frac{2\,m}{E-\varepsilon_{l\n}\n}}  \,  v_{l\n k}$ and $(A_i)_{lk}=(B)_{lk}$ for $k\neq i$ and for $k=i$ the column is replaced by a column with zeros in all the entries except entry $j$ which is equal to  $\frac{1}{(2\,\pi\,\hbar)^{1/2}}$. 

Next, we focus on the dependence of the determinants on $g$. Notice that the only dependence on the coupling strength $g$ is through $v_{l\n k}$, which are linear with $g$. The highest order in $g$ comes from the product of all possible $v_{l\n k}$ factors that appear in all the matrix elements. For $Det[B]$, the highest order term on $g$ is of order $N$. This factor can be calculated by ignoring the terms $\delta_{lk}$ in $B$, and it is proportional to the determinant of $\hat{S}_N$. In contrast, the maximum order on $g$ of $Det[A_i]$ is $N-1$. This is a consequence that the column $i$ does not depend on $g$ at all. This difference in power produces the $1/g$ behavior of the wavefunction at strong coupling.

In the weak-coupling limit, we focus on the lowest order in $g$. For $Det[B]$, it is of order zero and is produced by the product of the factors $\delta_{lk}$ which reside in the diagonal of $B$. For $Det[A_i]$, the lowest order is of order 1, because $(A_i)_{i,i}=0$, the determinant does not include any contribution exclusively from the diagonal elements. At least one off-diagonal element is required and is linear in $g$. The difference between the powers produces a linear behavior of the wavefunction for small $g$. This implies that for $g\rightarrow0$ the wavefunction inside the interaction region goes to zero (basically the interaction region disappears, because the interaction Hamiltonian becomes zero).

\section{Transition rates for the delta potentials model}

In this section, we derive the transition rates as a function of the wavefunction in the interaction region for a 1D delta potential (see equation 5 in the main text).

In the low-density limit, the transition rates are \cite{alicki_violation_2023}

\begin{gather}
a^k_{lj} = 2\nu_k \pi \int d\mathbf{p}\int d\mathbf{p}'\, Z_k^{-1}e^{-\beta_k}
\delta \bigl\{(E_{\mathbf{p}'} +\mathcal{E}_l)- ( E_{\mathbf{p}} +\mathcal{E}_{j} )\bigr\}|\langle l , \mathbf{p}'| T |\mathbf{p}, j\rangle|^2 .
\label{transprob}
\end{gather}

After integration over the delta function and changing variables, we get 

\begin{gather}
a^k_{lj} = 2\nu_k \pi m \int_{max \{\varepsilon_l,\varepsilon_j\}}^{\infty}dE Z^{-1}_k e^{-\beta_k(E-\varepsilon_j)} \frac{|\langle l , \mathbf{p}'| T |\mathbf{p}, j\rangle|^2}{\sqrt{E-\varepsilon_l} \sqrt{E-\varepsilon_j}}.
\label{transprob}
\end{gather}

Moreover, we can relate the $T-$matrix elements to the interaction Hamiltonian.

  \be \label{T-matels-take-0}
         \la {\bm p}'\n \,\n l\n | \,T \, | {\bm p} \,\n j \ra \; = \; \la {\bm p}'\n \,\n l\n | \, \hat{H}_{int} \, | \Psi \ra
      \ee

For $l\neq j$ we use \ref{coupling} and \ref{eq:linsys} to rewrite the  $T-$matrix elements as

 \be \label{T-matels-take-0}
         \la {\bm p}'\n \,\n l\n | \,T \, | {\bm p} \,\n j \ra = i \sqrt{\frac{\hbar(E-\varepsilon_l)}{m\pi}}
         \phi_{lj}(0).
      \ee


\section{Numerical simulations and figure parameters}

\subsection{System and common parameters}

We consider the following system Hamiltonian for an \(N\)-level system:
\begin{equation}
    \hat{H}_S
    =
    \tau
    \sum_{i=1}^{N}
    \left(
        e^{-i\theta_N}
        |\chi_i\rangle\langle \chi_{i+1}|
        +
        e^{i\theta_N}
        |\chi_{i+1}\rangle\langle \chi_i|
    \right),
    \qquad
    \theta_N=\frac{2\pi\phi}{N},
\end{equation}
with periodic boundary conditions,
\begin{equation}
    |\chi_{N+1}\rangle\equiv|\chi_1\rangle .
\end{equation}
Therefore, the term \(i=N\) describes the hopping between the last and first
sites arrange in a ring. The eigenvalues of \(\hat{H}_S\) are denoted by
\(\varepsilon_j\), and the corresponding eigenstates are \(|j\rangle\). The
incoming particle is a one-dimensional free particle with kinetic energy
\(E_p=p^2/(2m)\), so that
\begin{equation}
    \hat{H}_0|p\,j\rangle=(E_p+\varepsilon_j)|p\,j\rangle .
\end{equation}
Unless stated otherwise, the simulations use
\begin{equation}
    \tau=1,
    \qquad
    \phi=0.3,
    \qquad
    m=1,
    \qquad
    \hbar=1,
    \qquad
    k_B=1 .
\end{equation}
The bath parameters are
\begin{equation}
    \beta_h=1/2,
    \qquad
    \beta_c=1,
    \qquad
    \nu_h=\nu_c=1 .
\end{equation}

The coupling strength shown on the horizontal axis is the dimensionless
coupling
\begin{equation}
    \tilde g=\frac{g}{\omega},
    \qquad
    \omega=
    \max_j\varepsilon_j-\min_j\varepsilon_j .
\end{equation}
Here \(g\) denotes the coupling strength that enters the Hamiltonian, while
\(\tilde g\) is the dimensionless value shown in the plots.

When comparing an $N$-level system with an $(N+1)$-level system, the hopping
amplitude of the $(N+1)$-level system is chosen so that both systems have the
same value of $\omega$. For the comparison between $N=2$ and $N+1=3$,
with $\tau_2=1$ and $\phi_2=\phi_3=0.3$, this gives
\begin{equation}
    \tau_3\simeq 0.68245 .
\end{equation}

\subsection{Bump potential and coupling normalization}

We use the smooth bump potential
\begin{equation}
    b_a(x)=
    \begin{cases}
    \displaystyle
    \exp\left[-\frac{1}{a^2-x^2}\right],
    & |x|<a,\\[6pt]
    0,
    & |x|\ge a ,
    \end{cases}
\end{equation}
where \(a\) is the bump half-width. The support of the potential is therefore
\(x\in[-a,a]\). In the notation of the main text, the full width is \(L=2a\).
For the main bump figures we use
\begin{equation}
    a=1 .
\end{equation}

The  site coefficients are denoted by $c_i$. We normalize the full
site-dependent potential by
\begin{equation}
    v_i
    =
    \frac{c_i}
    {
    \left(\sum_r |c_r|\right)
    \displaystyle\int_{-a}^{a} b_a(x)\,dx
    } .
\end{equation}
Thus,
\begin{equation}
    \sum_i |v_i|
    \int_{-a}^{a} b_a(x)\,dx
    =
    1 .
\end{equation}
The site-dependent amplitude used in the scattering calculation is
\begin{equation}
    V_i(\tilde g)
    =
    gv_i
\end{equation}
  
In the energy eigenbasis of $\hat{H}_S$, the coupling matrix is
\begin{equation}
    v_{lk}(\tilde g)
    =
    \sum_i
    \langle l|\chi_i\rangle
    V_i(\tilde g)
    \langle \chi_i|k\rangle .
\end{equation}

For the delta-potential calculations, the integral of the delta function is
unity. The corresponding normalized weights are therefore
\begin{equation}
    v_i=\frac{c_i}{\sum_r |c_r|},
\end{equation}
and the site amplitude is
\begin{equation}
    V_i^{\delta}(\tilde g)=gv_i .
\end{equation}

\subsection{Finite-range scattering calculation}

The finite-range scattering calculation follows the multichannel
finite-difference scattering formulation of
Ref.~\cite[Sec.~G.4]{Jiene-thesis}, see also Ref.~\cite{Jiene-paper}. In that formulation, one
solves a stationary multichannel Schr\"odinger equation with a channel-space
matrix potential. In the present model, we generalize the fixed incoming
channel used there to an arbitrary incoming channel \(j\). The channel-space
potential matrix is
\begin{equation}
    {\cal V}_{lk}(x)
    =
    \varepsilon_l\delta_{lk}
    +
    b_a(x)v_{lk}(\tilde g).
\end{equation}
Therefore, for each incoming channel $j$ and scattering energy
\begin{equation}
    E=E_p+\varepsilon_j,
\end{equation}
we solve
\begin{equation}
    -\frac{\hbar^2}{2m}
    \frac{d^2\phi_{lj}(x;E)}{dx^2}
    +
    \sum_k
    {\cal V}_{lk}(x)\phi_{kj}(x;E)
    =
    E\phi_{lj}(x;E).
\end{equation}
Equivalently,
\begin{equation}
    -\frac{\hbar^2}{2m}
    \frac{d^2\phi_{lj}(x;E)}{dx^2}
    +
    \varepsilon_l\,\phi_{lj}(x;E)
    +
    b_a(x)
    \sum_k v_{lk}(\tilde g)\phi_{kj}(x;E)
    =
    E\,\phi_{lj}(x;E).
\end{equation}

The calculation is performed on the spatial grid
\begin{equation}
    x\in[-4,4],
    \qquad
    n_x=1601 .
\end{equation}
The second derivative is discretized by a central finite difference, and the
resulting linear boundary-value problem is solved with scattering boundary
conditions. 

For every open outgoing channel $l$, we define 
\begin{equation}
    k_l(E)=\frac{\sqrt{2m(E-\varepsilon_l)}}{\hbar}.
\end{equation}
For an incoming wave from the left in channel $j$, the asymptotic form is
\begin{equation}
    \phi_{lj}(x;E)
    =
    \frac{1}{(2\pi\hbar)^{1/2}}
    \left[
        \delta_{lj}e^{ik_jx}
        +
        \sqrt{\frac{k_j}{k_l}}\,
        {\cal R}_{lj}(E)e^{-ik_lx}
    \right],
    \qquad x<-a ,
\end{equation}
and
\begin{equation}
    \phi_{lj}(x;E)
    =
    \frac{1}{(2\pi\hbar)^{1/2}}
    \sqrt{\frac{k_j}{k_l}}\,
    {\cal T}_{lj}(E)e^{ik_lx},
    \qquad x>a .
\end{equation}
Here ${\cal T}_{lj}$ and ${\cal R}_{lj}$ are the flux-normalized transmission
and reflection amplitudes extracted from the numerical scattering solution.

Flux conservation was verified:
\begin{equation}
    \sum_{l\in\mathrm{open}}
    \left(
        |{\cal T}_{lj}(E)|^2
        +
        |{\cal R}_{lj}(E)|^2
    \right)
    \simeq 1 .
\end{equation}

\subsection{$T$-matrix elements}

The on-shell $T$-matrix is defined by
\begin{equation}
    \langle p'l|\hat T|pj\rangle
    =
    \langle p'l|V|\Psi_j^+(E)\rangle .
\end{equation}
The transmission and reflection amplitudes give
\begin{equation}
\label{eq:T_transmission_numerical}
    \langle p_l,l|\hat T|p_j,j\rangle
    =
    \frac{i}{2\pi m}
    \sqrt{p_lp_j}
    \left(
        {\cal T}_{lj}(E)-\delta_{lj}
    \right),
\end{equation}
and
\begin{equation}
\label{eq:T_reflection_numerical}
    \langle -p_l,l|\hat T|p_j,j\rangle
    =
    \frac{i}{2\pi m}
    \sqrt{p_lp_j}
    {\cal R}_{lj}(E).
\end{equation}
These are the on-shell $T$-matrix elements used in the finite-bump
transition-rate calculation. Closed outgoing channels do not contribute to the
final on-shell rates.

This is the finite-range analogue of the delta-potential expression derived
above: in the delta case the interaction region collapses to the single point
$x=0$, while in the bump case the same on-shell $T$-matrix information is
obtained numerically from the transmission and reflection amplitudes.

\subsection{Transition rates and heat current}

For the finite bump, the numerical scattering calculation is performed for
particles incident from the left. Since the potential is symmetric under
$x\rightarrow -x$, the contribution from incidence from the right is identical.
Therefore, for a transition $j\rightarrow l$, the squared matrix element used
in the rate integral is
\begin{equation}
\label{eq:Mlj_bump}
    M_{lj}(E)
    =
    2\left[
    \left|
    \langle p_l,l|\hat T|p_j,j\rangle
    \right|^2
    +
    \left|
    \langle -p_l,l|\hat T|p_j,j\rangle
    \right|^2
    \right],
\end{equation}
where the first term corresponds to transmission and the second to reflection.
The factor of two accounts for the identical contribution from incidence from
the opposite direction.

Thus, for the finite-bump calculation, the bath-induced transition rates are
computed as
\begin{equation}
\label{eq:finite_bump_rates}
    a_{lj}^{k}=
    2\pi m\nu_{k}
    \int_{\max(\varepsilon_l,\varepsilon_j)}^{\infty}
    dE\,
    Z_{k}^{-1}
    e^{-\beta_{k}(E-\varepsilon_j)}
    \frac{
    M_{lj}(E)
    }
    {
    \sqrt{E-\varepsilon_l}
    \sqrt{E-\varepsilon_j}
    },
    \qquad
    k=h,c .
\end{equation}

The total rate is
\begin{equation}
    a_{lj}=a_{lj}^{h}+a_{lj}^{c}.
\end{equation}
The steady-state populations are obtained from the Pauli master equation,
\begin{equation}
    \sum_j
    \left(
        a_{ij}p_j^{\mathrm{ss}}
        -
        a_{ji}p_i^{\mathrm{ss}}
    \right)
    =
    0,
    \qquad
    \sum_i p_i^{\mathrm{ss}}=1 .
\end{equation}
The heat current from bath $k$ is
\begin{equation}
    J^{k}
    =
    \sum_{i>j}
    (\varepsilon_i-\varepsilon_j)
    \left(
        a_{ij}^{k}p_j^{\mathrm{ss}}
        -
        a_{ji}^{k}p_i^{\mathrm{ss}}
    \right).
\end{equation}
In the figures we plot the heat current from the hot bath,
\begin{equation}
    J=J^h,
\end{equation}
with the steady-state condition giving
\begin{equation}
    J^h+J^c=0 .
\end{equation}

\subsection{Figure parameters}

\subsubsection*{Parameters for Fig.~1: bump heat-current curve}
The finite bump potential was used with half-width
\begin{equation}
    a=1,
\end{equation}
so that the interaction region is
\begin{equation}
    x\in[-1,1].
\end{equation}
The scattering calculation was performed on the spatial grid
\begin{equation}
    x\in[-4,4],
    \qquad
    n_x=1601 .
\end{equation}
For the two-level system, we used
\begin{equation}
    N=2,
    \qquad
    (c_1,c_2)=(1.8,0.5),
\end{equation}
where the coefficients \(c_i\) were normalized according to the prescription
given above. The heat current was computed as a function of the normalized
coupling
\begin{equation}
    \tilde g=\frac{g}{\omega} .
\end{equation}
For each value of \(\tilde g\), the scattering problem was solved, the
transition rates were computed, the steady state was obtained, and the heat
current \(J_h\) was evaluated.

\subsubsection*{Parameters for Fig.~2: Wavefunction-weight figure}
For Fig.~2, we used the same finite bump potential and spatial grid as in
Fig.~1,
\begin{equation}
    a=1,
    \qquad
    x\in[-4,4],
    \qquad
    n_x=1601 .
\end{equation}
The site coefficients used for the two-level system were
\begin{equation}
    N=2,
    \qquad
    (c_1,c_2)=(1.8,0.5),
\end{equation}
with the same bump normalization prescription described above.

For each value of the normalized coupling \(\tilde g\), the finite-range
scattering problem was solved at the fixed probe energy
\begin{equation}
    E_{\mathrm{probe}}
    =
    \max_l\varepsilon_l+1/\beta_h .
\end{equation}
This choice ensures that all system channels are open. We then computed the
normalized wavefunction weight inside the interaction region,
\begin{equation}
    W_a(\tilde g)
    =
    \frac{1}{2a}
    \sum_{j\in\mathrm{open}}
    \int_{-a}^{a}
    dx
    \sum_l
    |\phi_{lj}(x;E_{\mathrm{probe}},\tilde g)|^2 .
\end{equation}
Here \(j\) labels the incoming open channel and \(l\) labels the wavefunction
component. The factor \(1/(2a)\) normalizes by the width of the interaction
region. Since \(a=1\), this is equivalent to dividing the raw integral by \(2\).

The plotted curve shows \(W_a(\tilde g)\) as a function of
\(\tilde g=g/\omega\).

\subsubsection*{Parameters for Fig.~3: delta-potential \(N\) versus \(N+1\) comparison}

For Fig.~3, we used the analytic delta-potential calculation described above.

The two-level system had
\begin{equation}
    N=2,
    \qquad
    (c_1,c_2)=(1.8,0.2),
\end{equation}
with normalized delta-potential weights
\begin{equation}
    v_i=\frac{c_i}{\sum_r |c_r|}.
\end{equation}

The three-level system included an additional uncoupled level,
\begin{equation}
    N+1=3,
    \qquad
    (c_1,c_2,c_3)=(1.8,0.2,0).
\end{equation}

All other system and bath parameters were the common parameters listed above.
The hopping parameter of the three-level system was chosen so that the
two-level and three-level systems had the same value of \(\omega\),
as described above. The heat current was computed as a function of the
normalized coupling
\begin{equation}
    \tilde g=\frac{g}{\omega} .
\end{equation}

\subsubsection*{Parameters for Fig.~4: delta-potential ratio figure}

For Fig.~4, we used the analytic delta-potential calculation described above.
All other system and bath parameters were the common parameters listed above.

Following the notation used in the figure, we varied the normalized relative
delta-potential weight \(v_1\) in the range \(0.01\le v_1\le0.99\). The
remaining weights were chosen such that \(v_1+v_2=1\), with \(v_3=0\). Since
\(v_1+v_2\) was fixed throughout the scan, the horizontal axis \(v_1-v_2\) is
equivalent to the normalized asymmetry \((v_1-v_2)/(v_1+v_2)\).

For each value of the asymmetry, we computed the two-level maximum current
\begin{equation}
   Max[J^h_2(g)]
    =
    \max_{ g} J^h_{2}( g).
\end{equation}

For the three-level system, the large-coupling current was evaluated at
\begin{equation}
   g=1000,
\end{equation}
and denoted by
\begin{equation}
    Max[J^h_3(g)]
\end{equation}
The plotted quantity was
\begin{equation}
    R
    =
    \frac{Max[J^h_3(g)]}{Max[J^h_2(g)]} .
\end{equation}

The ratio was plotted on a logarithmic \(y\)-scale. For visualization only,
very large values of the ratio near the symmetric point were capped at
\(R\sim10^3\).

\section{Scattering in the case of infinitely repulsive, localized, abrupt one-dimensional potentials:
No penetration into the infinitely repulsive region, total reflection, no transitions between channels}

In this last Section of the Supplementary Information, we provide a detailed mathematical proof of the
"no transitions \& total reflection" claim which is exploited in the main text. For the sake of simplicity,
we restrict ourselves here to a broad class of localized and abrupt interaction potentials in one spatial dimension.

\subsection{One channel setup (as a motivation and a warmup)}

      \textbf{\textit{Preliminaries.}}\\
      Consider single channel scattering in 1D, with a finite valued potential $V\m(x)$\\
      which vanishes outside $(a,b)$ and which is continuous inside $(a,b)$.\\
      Our scattering problem is formulated mathematically as follows:\\
      The time independent Schr\"{o}dinger equation (TISE)
      \be \label{TISE-1D}
         -\,\frac{\hbar^2}{2\,m} \, \frac{\partial^2}{\partial x^2} \, \psi(x) \; + \; V\m(x) \, \psi(x) \; = \; E \, \psi(x) \mez ;
      \ee
      with the boundary conditions
      \be \label{bc-1D}
         \psi(x \leq a) \; = \; e^{+ikx} \; + \; R(E) \, e^{-ikx} \mez , \mez
         \psi(x \geq b) \; = \; T(E) \, e^{+ikx} \mez ;
      \ee
      here of course $\hbar\,k=\sqrt{2\,m\,E}$.\\
      The boundary value problem (\ref{TISE-1D})-(\ref{bc-1D}) possesses always an unique solution\\
      with $T(E)$ being always nonzero: {\it
      An initial value problem
      $$
        -\,\frac{\hbar^2}{2\,m} \, \frac{\partial^2}{\partial x^2} \, \tilde{\psi}(x) \; + \; V\m(x) \, \tilde{\psi}(x)
        \; = \; E \, \tilde{\psi}(x) \mez , \mez \tilde{\psi}(x \ge b) \; = \; e^{+ikx}
      $$
      possesses an unique solution $\tilde{\psi}(x)$ defined for all $x \in {\mathbb R}$. One has
      $$
        \tilde{\psi}(x \le a) \; = \; A(E) \, e^{+ikx} \; + \; B(E) \, e^{-ikx} \mez ;
      $$
      where $A(E) \neq 0$ (having $A(E)=0$ would violate flux conservation).\\
      The solution $\psi(x)$ of (\ref{TISE-1D})-(\ref{bc-1D}) is obtained by setting $\psi(x)=\tilde{\psi}(x)/A(E)$.}\\
      We have
      \be \label{psi-a-b}
         \psi(a) \; = \; e^{+ika} \; + \; R(E) \, e^{-ika} \mez , \mez
         \psi(b) \; = \; T(E) \, e^{+ikb} \mez ;
      \ee
      and
      \be \label{psi-prime-a-b}
         \psi'\n(a) \; = \; i \, k \, e^{+ika} \; - \; i \, k \, R(E) \, e^{-ika} \mez , \mez
         \psi'\n(b) \; = \; i \, k \, T(E) \, e^{+ikb} \mez .
      \ee
      Recall also the probability conservation property
      \be \label{prb-1D}
         \Bigl|T(E)\Bigr|^2 \, + \, \Bigl|R(E)\Bigr|^2 \, = \, 1 \mez .
      \ee
      Direct calculation utilizing (\ref{psi-a-b}), (\ref{psi-prime-a-b}), (\ref{prb-1D}) yields
      \be \label{entity-1D}
         \psi^*\n(a) \, \psi'\n(a) \, - \, \psi^*\n(b) \, \psi'\n(b) \; = \;
         i\,k \, \Bigl( R^*\m(E) \, e^{+2ika} \, - \, R(E) \, e^{-2ika} \Bigr) \mez ;
      \ee
      this is apparently a real valued entity.\\
      Recall also the Wronskian
      \be \label{W}
         W \; = \; \psi^*\n(x) \, \psi'\n(x) \; - \; \psi'^*\n(x) \, \psi(x) \mez .
      \ee
      This Wronskian is independent upon $x$\\ (as demonstrated in any textbook on ordinary differential equations).\\
      Note that $W$ is closely related to the flux (as demonstrated in any quantum mechanics textbook).\\
      The Wronskian property
      \be
         \psi^*\n(a) \, \psi'\n(a) \; - \; \psi'^*\n(a) \, \psi(a) \; = \; \psi^*\n(b) \, \psi'\n(b) \; - \; \psi'^*\n(b) \, \psi(b)
      \ee
      can be redisplayed in an equivalent appearance
      \be
         \psi^*\n(a) \, \psi'\n(a) \; - \; \psi^*\n(b) \, \psi'\n(b) \; = \; \psi(a) \, \psi'^*\n(a) \; - \; \psi(b) \, \psi'^*\n(b) \mez .
      \ee
      This shows again that entity (\ref{entity-1D}) must be real.\\
      \textbf{\textit{Auxiliary elaborations.}}\\
      Take (\ref{TISE-1D}), multiply by $\psi^*\n(x)$, and integrate over $x \in (a,b)$. One gets
      \be
         -\,\frac{\hbar^2}{2\,m} \, \int_{a}^{b} \m {\rm d}x \; \psi^*\n(x) \, \psi''\n(x) \; + \;
         \int_{a}^{b} \m {\rm d}x \; \psi^*\n(x) \, V\m(x) \, \psi(x) \; = \; E \, \int_{a}^{b} \m {\rm d}x \; \psi^*\n(x) \, \psi(x) \mez .
      \ee
      Integration by parts yields subsequently
      \begin{eqnarray} \label{E-int-relation-1D-take-1}
         & & \frac{\hbar^2\,\n i\,k}{2\,m} \, \Bigl( R^*\m(E) \, e^{+2ika} \, - \, R(E) \, e^{-2ika} \Bigr) \\
         & + & \frac{\hbar^2}{2\,m} \, \int_{a}^{b} \m {\rm d}x \; \psi'^*\n(x) \, \psi'\n(x) \; + \;
         \int_{a}^{b} \m {\rm d}x \; \psi^*\n(x) \, V\m(x) \, \psi(x) \; = \;
         E \, \int_{a}^{b} \m {\rm d}x \; \psi^*\n(x) \, \psi(x) \mez ; \nonumber
      \end{eqnarray}
      we have utilized here (\ref{entity-1D}). The integral
      \be \label{integral-1D}
         I_0 \; = \; \int_{a}^{b} \m {\rm d}x \; \psi^*\n(x) \, \psi(x)
      \ee
      is inevitably nonzero. Having $I_0=0$ would mean that $\psi(x)$ vanishes identically in $(a,b)$.\\
      Instead of (\ref{E-int-relation-1D-take-1}) we can thus write alternatively
      \begin{eqnarray} \label{E-1D-take-2}
         \hspace*{-1.00cm} E & = & \frac{\hbar^2\,\n i\,k}{2\,m} \,
         \frac{\Bigl( R^*\m(E) \, e^{+2ika} \, - \, R(E) \, e^{-2ika} \Bigr)}{\int_{a}^{b} \m {\rm d}x \; \psi^*\n(x) \, \psi(x)}
         \; + \; \frac{\hbar^2}{2\,m} \, \frac{\int_{a}^{b} \m {\rm d}x \; \psi'^*\n(x) \, \psi'\n(x)}
         {\int_{a}^{b} \m {\rm d}x \; \psi^*\n(x) \, \psi(x)} \; + \;
         \frac{\int_{a}^{b} \m {\rm d}x \; \psi^*\n(x) \, V\m(x) \, \psi(x)}
         {\int_{a}^{b} \m {\rm d}x \; \psi^*\n(x) \, \psi(x)} \mz . \mz
      \end{eqnarray}
      \textbf{\textit{Restricting further the potentials \& scaling by $\bm{g \to \infty}$.}}\\
      Suppose hereafter that the potential $V\m(x)$ satisfies the positivity condition
      \be
         0 \, < \, V_{\min} \, < \, V\m(x) \mez . \mez (\forall\,x)(a<x<b)
      \ee
      This also means that $V\m(x)$ is discontinuous at the boundaries $x=a$ and $x=b$.\\
      Suppose also that $V\m(x)$ is scaled by $g \to +\infty$,\\
      such that $g\,V\m(x)$ uniformly approaches $+\infty$ for $x\in(a,b)$.\\
      Equation (\ref{E-int-relation-1D-take-1}) is slightly updated into
      \begin{eqnarray} \label{E-int-relation-1D-take-1-updated}
         & & \frac{\hbar^2\,\n i\,k}{2\,m} \, \Bigl( R^*\m(E) \, e^{+2ika} \, - \, R(E) \, e^{-2ika} \Bigr) \\
         & + & \frac{\hbar^2}{2\,m} \, \int_{a}^{b} \m {\rm d}x \; \psi'^*\n(x) \, \psi'\n(x) \; + \;
         \int_{a}^{b} \m {\rm d}x \; \psi^*\n(x) \, g\,V\m(x) \, \psi(x) \; = \;
         E \, \int_{a}^{b} \m {\rm d}x \; \psi^*\n(x) \, \psi(x) \mez . \nonumber
      \end{eqnarray}
      Equation (\ref{E-1D-take-2}) is slightly updated into
      \begin{eqnarray} \label{E-1D-take-3}
         \hspace*{-1.25cm} E & = & \frac{\hbar^2\,\n i\,k}{2\,m} \,
         \frac{\Bigl( R^*\m(E) \, e^{+2ika} \, - \, R(E) \, e^{-2ika} \Bigr)}{\int_{a}^{b} \m {\rm d}x \; \psi^*\n(x) \, \psi(x)}
         \; + \; \frac{\hbar^2}{2\,m} \, \frac{\int_{a}^{b} \m {\rm d}x \; \psi'^*\n(x) \, \psi'\n(x)}
         {\int_{a}^{b} \m {\rm d}x \; \psi^*\n(x) \, \psi(x)} \; + \;
         \frac{\int_{a}^{b} \m {\rm d}x \; \psi^*\n(x) \, g\,V\m(x) \, \psi(x)}
         {\int_{a}^{b} \m {\rm d}x \; \psi^*\n(x) \, \psi(x)} \mz . \mz
      \end{eqnarray}
      Here both $\psi(x)$ and $R(E)$ depend also upon $g$.\\
      \textbf{\textit{Insight \#1.}}\\
      We have
      \be
         0 \; < \; g\,V_{\min} \; < \;
         \frac{\int_{a}^{b} \m {\rm d}x \; \psi^*\n(x) \, g\,V\m(x) \, \psi(x)}{\int_{a}^{b} \m {\rm d}x \; \psi^*\n(x) \, \psi(x)} \mez .
      \ee
      Thus the term
      \be
         \frac{\int_{a}^{b} \m {\rm d}x \; \psi^*\n(x) \, g\,V\m(x) \, \psi(x)}{\int_{a}^{b} \m {\rm d}x \; \psi^*\n(x) \, \psi(x)}
      \ee
      inevitably diverges to $+\infty$ as $g \to +\infty$. Moreover, we have
      \be
         0 \; \le \; \frac{\int_{a}^{b} \m {\rm d}x \; \psi'^*\n(x) \, \psi'\n(x)}{\int_{a}^{b} \m {\rm d}x \; \psi^*\n(x) \, \psi(x)} \mez .
      \ee
      Returning back to equation (\ref{E-1D-take-3}), we infer that the term
      \be
         \frac{\hbar^2\,\n i\,k}{2\,m} \,
         \frac{\Bigl( R^*\m(E) \, e^{+2ika} \, - \, R(E) \, e^{-2ika} \Bigr)}{\int_{a}^{b} \m {\rm d}x \; \psi^*\n(x) \, \psi(x)}
      \ee
      must inevitably diverge to $-\infty$ as $g \to +\infty$.\\
      If so, then the integral $I_0=(\ref{integral-1D})$ must inevitably approach zero as $g \to +\infty$.\\
      {\sl In other words, there is no penetration into the strongly repulsive interaction region.}\\
      \textbf{\textit{Insight \#2.}}\\
      Consider now equation (\ref{E-int-relation-1D-take-1-updated}).\\
      As $g \to +\infty$, the r.h.s.~of (\ref{E-int-relation-1D-take-1-updated}) approaches zero.\\
      Hence none of the two positive valued integrals
      \be \label{positive-integrals}
         \int_{a}^{b} \m {\rm d}x \; \psi'^*\n(x) \, \psi'\n(x) \mez , \mez
         \int_{a}^{b} \m {\rm d}x \; \psi^*\n(x) \, g\,V\m(x) \, \psi(x)
      \ee
      can diverge to $+\infty$ in the limit of $g \to +\infty$.\\
      Each of these integrals can either stay finite valued or go to zero.\\
      \textbf{\textit{Insight \#3.}}\\
      Recall that $I_0=(\ref{integral-1D})$ approaches zero as $g \to +\infty$,\\
      while the integral $\int_{a}^{b} \m {\rm d}x \; \psi'^*\n(x) \, \psi'\n(x)$ remains bounded (non-divergent).\\
      This means that $\psi(x)$ falls off to zero almost everywhere in $(a,b)$,\\
      while $\psi'\n(x)$ remains bounded in magnitude almost everywhere in $(a,b)$.\\
      If so, then the Wronskian (\ref{W}) inevitably vanishes in the limit of $g \to +\infty$.\\
      \textbf{\textit{Insight \#4.}}\\
      If the Wronskian $W=(\ref{W})\,=\,2\,i\,k\,|T(E)|^2$ vanishes in the limit of $g \to +\infty$,\\
      then the same must be true also for $T(E)$ at $g \to +\infty$.\\
      Hence also the modulus $|R(E)|_{g \to +\infty}=1$.\\
      {\sl Total reflection occurs in the limit of $g \to +\infty$.}\\
      \textbf{\textit{Insight \#5.}}\\
      For $g \to +\infty$ we thus have $R(E)=e^{i\phi(E)}$\n,\\
      where $\phi(E)$ is an as yet unknown phase (possibly even $g$-dependent).\\
      Thus $\psi(a) \, = \, e^{+ika} \, + \, e^{-i(ka-\phi(E))}$\m, and $\psi(a<x<b)=0$ almost everywhere.\\
      We claim now that $\phi(E)\,=\,2\,k\,a\,+\,\pi$ such that $R(E)=-e^{+ika}$ and $\psi(a)=0$.\\
      The justification is as follows: Suppose that $R(E)\neq-e^{+ika}$\m,\\
      then $\psi'\n(x)$ would for $g \to +\infty$ exhibit a $\delta$-type jump as $x$ approaches $a$ from the right.\\
      Making the integral
      \be
         \int_{a}^{b} \m {\rm d}x \; \psi'^*\n(x) \, \psi'\n(x)
      \ee
      divergent to $+\infty$ as $g \to +\infty$. We have however excluded this possibility above in (\ref{positive-integrals}).\\
      {\sl Thus, for $g \to +\infty$, the wavefunction $\psi(x)$ reflects at $x=a$ as a standing wave (node at $x=a$).}

     \subsection{Multichannel setup (relevant to the scattering problems considered in the main text)}

      \textbf{\textit{Introducing the problem.}}\\
      The total scattering wavefunction is expanded as
      \be
         | \Psi \ra \; = \; \sum_{j=1}^{N} \Psi_{j}\n(x) \, |j\ra \mez ;
      \ee
      where $|j\ra$ are the eigenstates of $\hat{H}_{\rm S}$.\\
      Exactly as done in the main text (just with the notations a bit altered).\\
      To simplify our subsequent considerations,\\ we switch into matrix language and introduce accordingly a shorthand symbol
      \be
         \bm{\Psi}(x) \; = \; \left( \begin{matrix} \Psi_{1}\n(x) \cr \bm\vdots \cr \Psi_{N}\n(x) \end{matrix} \right) \mez .
      \ee
      By construction, $\bm{\Psi}(x)$ satisfies the TISE
      \be \label{Psi-CC-eigenproblem}
         -\,\frac{\hbar^2}{2\,m} \, \partial_{xx} \, \bm{\Psi}(x) \; + \; \bm{W}\m(x) \, \bm{\Psi}(x) \; = \; E \, \bm{\Psi}(x) \mez ;
      \ee
      with
      \be
         \bm{W}\m(x) \; = \; \bm{F} \; + \; \bm{V}\m(x) \mez ;
      \ee
      and obvious notations
      \be
         \bm{F} \; = \; {\rm diag}\{ \varepsilon_j \} \mez , \mez
         (\bm{V}\m(x))_{jj'} \; = \; \sum_{i=1}^N \, \la j | \chi_i \ra \, g\,V_i\n(x) \, \la \chi_i | j' \ra \mez .
      \ee
      The matrix $\bm{V}\m(x)$ vanishes outside $(a,b)$ and depends continuously upon $x$ inside $(a,b)$.\\
      Scattering boundary conditions are imposed,
      \begin{eqnarray}
         \label{Psi-bc-left} \Psi_j\n(x \le a) & = &
         \delta_{jj_{\rm in}} \; e^{+ik_{j_{\rm in}}x} \; + \; \sqrt{\frac{k_{j_{\rm in}}}{k_j}} \; R_j\n(E) \; e^{-i k_j x} \mez ;\\
         \label{Psi-bc-right} \Psi_j\n(x \ge b) & = & \hspace*{+2.85cm} \sqrt{\frac{k_{j_{\rm in}}}{k_j}} \; T_j\n(E) \; e^{+i k_j x} \mez ;
      \end{eqnarray}
      here
      \be
         \hbar \, k_j \; = \; \sqrt{2\,m\,(E-\varepsilon_j)} \mez .
      \ee
      Note that $k_j$ is real positive for open channels,\\
      and positive imaginary for closed channels ($k_j=i\kappa_j$, with $\kappa_j>0$).\\
      The boundary value problem (\ref{Psi-CC-eigenproblem}) \& (\ref{Psi-bc-left})-(\ref{Psi-bc-right}) possesses always an unique solution.\\
      Recall also the probability conservation property
      \be \label{prb-CC}
         \sum_{j}^{E>\varepsilon_j} \, \Bigl|T_j\n(E)\Bigr|^2 \, + \, \Bigl|R_j\n(E)\Bigr|^2 \, = \, 1 \mez .
      \ee
      Direct calculation utilizing (\ref{Psi-bc-left}), (\ref{Psi-bc-right}), (\ref{prb-CC}) yields
      \begin{eqnarray} \label{Psi-term-take-1}
         \bm{\Psi}^\dagger\n(a) \; \bm{\Psi}'\n(a) \; - \; \bm{\Psi}^\dagger\n(b) \; \bm{\Psi}'\n(b) & = & k_{j_{\rm in}} \,
         i \, \Bigl( R_{j_{\rm in}}^*\n(E) \; e^{+2ik_{\rm in}a} \, - \, R_{j_{\rm in}}\n(E) \; e^{-2ik_{\rm in}a} \Bigr) \\
         & + & k_{j_{\rm in}} \, \sum_{j}^{E<\varepsilon_j}
         \left( \, \Bigl|R_{j}\n(E)\Bigr|^2 e^{+2\kappa_ja} \, + \, \Bigl|T_{j}\n(E)\Bigr|^2 e^{-2\kappa_jb} \right) \mez . \nonumber
      \end{eqnarray}
      This is apparently a real valued quantity.\\
      Recall also the Wronskian
      \be \label{W-CC}
         W \; = \; \bm{\Psi}^\dagger\n(x) \, \bm{\Psi}'\n(x) \; - \; \bm{\Psi}'^{\dagger}\n(x) \, \bm{\Psi}(x) \mez .
      \ee
      This Wronskian is independent upon $x$. Note that $W$ is closely related to the flux.\\
      The Wronskian property
      \be
         \bm{\Psi}^\dagger\n(a) \, \bm{\Psi}'\n(a) \; - \; \bm{\Psi}'^{\dagger}\n(a) \, \bm{\Psi}(a) \; = \;
         \bm{\Psi}^\dagger\n(b) \, \bm{\Psi}'\n(b) \; - \; \bm{\Psi}'^{\dagger}\n(b) \, \bm{\Psi}(b)
      \ee
      can be redisplayed in an equivalent appearance
      \be
         \bm{\Psi}^\dagger\n(a) \, \bm{\Psi}'\n(a) \; - \; \bm{\Psi}^\dagger\n(b) \, \bm{\Psi}'\n(b) \; = \;
         \bm{\Psi}^{T}\m(a) \, \bm{\Psi}'^{*}\n(a) \; - \; \bm{\Psi}^{T}\m(b) \, \bm{\Psi}'^{*}\n(b) \mez .
      \ee
      Showing again that entity (\ref{Psi-term-take-1}) must be real.\\
      \textbf{\textit{The transformation.}}\\
      We set
      \be
         U_{j'\iota'} \; = \; \la j'\n | \chi_{\iota'}\n \ra \mez ;
      \ee
      and define
      \be
         \bm{\Phi}(x) \; = \; \bm{U} \, \bm{\Psi}(x) \mez .
      \ee
      One has then
      \be \label{Phi-CC-eigenproblem}
         -\,\frac{\hbar^2}{2\,m} \, \partial_{xx} \, \bm{\Phi}(x) \; + \; \bm{\widetilde{W}}\m(x) \, \bm{\Phi}(x) \; = \;
         E \, \bm{\Phi}(x) \mez ;
      \ee
      with
      \be
         \bm{\widetilde{W}}\m(x) \; = \; \bm{U}^\dagger \, \bm{W}\m(x) \, \bm{U} \; = \;
         \bm{\widetilde{F}} \; + \; \bm{\widetilde{V}}\m(x) \mez .
      \ee
      Here
      \be
         \bm{\widetilde{V}}\m(x) \; \equiv \; \bm{U}^\dagger \, \bm{V}\m(x) \, \bm{U} \; = \; {\rm diag}\{ g \, V_\iota\n(x) \} \mez ;
      \ee
      and
      \be
         \bm{\widetilde{F}} \; = \; \bm{U}^\dagger \bm{F} \, \bm{U} \mez .
      \ee
      Note that $\bm{\widetilde{F}}$ is an off-diagonal constant matrix.\\
      Instead of (\ref{Psi-term-take-1}) we can write equivalently also
      \begin{eqnarray} \label{Psi-term-take-2}
         \bm{\Phi}^\dagger\n(a) \; \bm{\Phi}'\n(a) \; - \; \bm{\Phi}^\dagger\n(b) \; \bm{\Phi}'\n(b) & = & k_{j_{\rm in}} \,
         i \, \Bigl( R_{j_{\rm in}}^*\n(E) \; e^{+2ik_{\rm in}a} \, - \, R_{j_{\rm in}}\n(E) \; e^{-2ik_{\rm in}a} \Bigr) \\
         & + & k_{j_{\rm in}} \, \sum_{j}^{E<\varepsilon_j}
         \left( \, \Bigl|R_{j}\n(E)\Bigr|^2 e^{+2\kappa_ja} \, + \, \Bigl|T_{j}\n(E)\Bigr|^2 e^{-2\kappa_jb} \right) \mez . \nonumber
      \end{eqnarray}
      Instead of (\ref{W-CC}) we can write equivalently also
      \be \label{W-CC-take-2}
         W \; = \; \bm{\Phi}^\dagger\n(x) \, \bm{\Phi}'\n(x) \; - \; \bm{\Phi}'^{\dagger}\n(x) \, \bm{\Phi}(x) \mez .
      \ee
      \textbf{\textit{Auxiliary elaborations.}}\\
      Take (\ref{Phi-CC-eigenproblem}), multiply by $\bm{\Phi}^\dagger\n(x)$, and integrate over $x \in (a,b)$. One gets
      \be
         -\,\frac{\hbar^2}{2\,m} \, \int_{a}^{b} \m {\rm d}x \; \bm{\Phi}^\dagger\n(x) \, \partial_{xx} \, \bm{\Phi}(x) \; + \;
         \int_{a}^{b} \m {\rm d}x \; \bm{\Phi}^\dagger\n(x) \, \bm{\widetilde{W}}\m(x) \, \bm{\Phi}(x) \; = \;
         E \, \int_{a}^{b} \m {\rm d}x \; \bm{\Phi}^\dagger\n(x) \, \bm{\Phi}(x) \mez .
      \ee
      Integration by parts yields subsequently
      \begin{eqnarray} \label{E-formula-CC-take-1}
         & & \frac{\hbar^2\,k_{j_{\rm in}}}{2\,m} \;
         i \, \Bigl( R_{j_{\rm in}}^*\n(E) \; e^{+2ik_{\rm in}a} \, - \, R_{j_{\rm in}}\n(E) \; e^{-2ik_{\rm in}a} \Bigr) \nonumber\\
         & + & \frac{\hbar^2\,k_{j_{\rm in}}}{2\,m} \, \sum_{j}^{E<\varepsilon_j}
         \left( \, \Bigl|R_{j}\n(E)\Bigr|^2 e^{+2\kappa_ja} \, + \, \Bigl|T_{j}\n(E)\Bigr|^2 e^{-2\kappa_jb} \right) \nonumber\\
         & + & \frac{\hbar^2}{2\,m} \, \int_{a}^{b} \m {\rm d}x \; \bm{\Phi}'^\dagger\n(x) \; \bm{\Phi}'\n(x) \; + \;
         \int_{a}^{b} \m {\rm d}x \; \bm{\Phi}^\dagger\n(x) \, \bm{\widetilde{W}}\m(x) \, \bm{\Phi}(x) \; = \\
         & = & E \, \int_{a}^{b} \m {\rm d}x \; \bm{\Phi}^\dagger\n(x) \, \bm{\Phi}(x) \mez . \nonumber
      \end{eqnarray}
      We have exploited here the property (\ref{Psi-term-take-2}).
      The integral
      \be \label{integral-CC}
         I_0 \; = \; \int_{a}^{b} \m {\rm d}x \; \bm{\Phi}^\dagger\n(x) \, \bm{\Phi}(x)
      \ee
      is inevitably nonzero. Having $I_0=0$ would mean that $\bm{\Phi}(x)$ vanishes identically in $(a,b)$.\\
      Instead of (\ref{E-formula-CC-take-1}) we can write more conveniently
      \begin{eqnarray} \label{E-formula-CC-take-2}
         E & = &
         \frac{\hbar^2\,k_{j_{\rm in}}}{2\,m} \left( \, \int_{a}^{b} \m {\rm d}x \; \bm{\Phi}^\dagger\n(x) \, \bm{\Phi}(x)\right)^{\m\m\m-1}
         i \, \Bigl( R_{j_{\rm in}}^*\n(E) \; e^{+2ik_{\rm in}a} \, - \, R_{j_{\rm in}}\n(E) \; e^{-2ik_{\rm in}a} \Bigr) \nonumber\\
         & + & \frac{\hbar^2\,k_{j_{\rm in}}}{2\,m}
         \left( \, \int_{a}^{b} \m {\rm d}x \; \bm{\Phi}^\dagger\n(x) \, \bm{\Phi}(x)\right)^{\m\m\m-1} \; \sum_{j}^{E<\varepsilon_j}
         \left( \, \Bigl|R_{j}\n(E)\Bigr|^2 e^{+2\kappa_ja} \, + \, \Bigl|T_{j}\n(E)\Bigr|^2 e^{-2\kappa_jb} \right) \nonumber\\
         & + & \frac{\hbar^2}{2\,m} \, \frac{\int_{a}^{b} \m {\rm d}x \; \bm{\Phi}'^\dagger\n(x) \, \bm{\Phi}'\n(x)}
         {\int_{a}^{b} \m {\rm d}x \; \bm{\Phi}^\dagger\n(x) \, \bm{\Phi}(x)} \; + \;
         \frac{\int_{a}^{b} \m {\rm d}x \; \bm{\Phi}^\dagger\n(x) \, \bm{\widetilde{V}}\m(x) \, \bm{\Phi}(x)}
         {\int_{a}^{b} \m {\rm d}x \; \bm{\Phi}^\dagger\n(x) \, \bm{\Phi}(x)} \; + \;
         \frac{\int_{a}^{b} \m {\rm d}x \; \bm{\Phi}^\dagger\n(x) \, \bm{\widetilde{F}} \, \bm{\Phi}(x)}
         {\int_{a}^{b} \m {\rm d}x \; \bm{\Phi}^\dagger\n(x) \, \bm{\Phi}(x)} \mez .
      \end{eqnarray}
      \textbf{\textit{Restricting further the potentials.}}\\
      Suppose hereafter that all the potentials $V_\iota\n(x)$ satisfy the positivity condition
      \be
         0 \, < \, V_{\min} \, < \, V_\iota\n(x) \mez . \mez (\forall\,x)(a<x<b)
      \ee
      This also means that $V_\iota\n(x)$ is discontinuous at the boundaries $x=a$ and $x=b$.\\
      \textbf{\textit{Insight \#1.}}\\
      We have
      \be
         0 \; < \; g\,v_{\min} \; < \;
         \frac{\int_{a}^{b} \m {\rm d}x \; \bm{\Phi}^\dagger\n(x) \, \bm{\widetilde{V}}\m(x) \, \bm{\Phi}(x)}
         {\int_{a}^{b} \m {\rm d}x \; \bm{\Phi}^\dagger\n(x) \, \bm{\Phi}(x)} \mez .
      \ee
      Thus the term
      \be
         \frac{\int_{a}^{b} \m {\rm d}x \; \bm{\Phi}^\dagger\n(x) \, \bm{\widetilde{V}}\m(x) \, \bm{\Phi}(x)}
         {\int_{a}^{b} \m {\rm d}x \; \bm{\Phi}^\dagger\n(x) \, \bm{\Phi}(x)}
      \ee
      inevitably diverges to $+\infty$ as $g \to +\infty$. Moreover, we have
      \be
         0 \; \le \; \frac{\int_{a}^{b} \m {\rm d}x \; \bm{\Phi}'^\dagger\n(x) \, \bm{\Phi}'\n(x)}
         {\int_{a}^{b} \m {\rm d}x \; \bm{\Phi}^\dagger\n(x) \, \bm{\Phi}(x)} \mez .
      \ee
      In addition, the term
      \be
         \frac{\int_{a}^{b} \m {\rm d}x \; \bm{\Phi}^\dagger\n(x) \, \bm{\widetilde{F}} \, \bm{\Phi}(x)}
         {\int_{a}^{b} \m {\rm d}x \; \bm{\Phi}^\dagger\n(x) \, \bm{\Phi}(x)}
      \ee
      is bounded in magnitude (its absolute value cannot exceed $\max\{|\varepsilon_j|\}_j$).\\
      Returning back to equation (\ref{E-formula-CC-take-2}), we infer that the term
      \be
         \frac{\hbar^2\,k_{j_{\rm in}}}{2\,m} \left( \, \int_{a}^{b} \m {\rm d}x \; \bm{\Phi}^\dagger\n(x) \, \bm{\Phi}(x)\right)^{\m\m\m-1}
         i \, \Bigl( R_{j_{\rm in}}^*\n(E) \; e^{+2ik_{\rm in}a} \, - \, R_{j_{\rm in}}\n(E) \; e^{-2ik_{\rm in}a} \Bigr)
      \ee
      must inevitably diverge to $-\infty$ as $g \to +\infty$.\\
      If so, then the integral $I_0=(\ref{integral-CC})$ must inevitably approach zero as $g \to +\infty$.\\
      {\sl In other words, there is no penetration into the strongly repulsive interaction region.}\\
      \textbf{\textit{Insight \#2.}}\\
      Consider now equation (\ref{E-formula-CC-take-1}).\\
      As $g \to +\infty$, the r.h.s.~of (\ref{E-formula-CC-take-1}) approaches zero.\\
      Hence none of the two positive valued integrals
      \be \label{positive-integrals-CC}
         \int_{a}^{b} \m {\rm d}x \; \bm{\Phi}'^\dagger\n(x) \; \bm{\Phi}'\n(x) \mez , \mez
         \int_{a}^{b} \m {\rm d}x \; \bm{\Phi}^\dagger\n(x) \, \bm{\widetilde{W}}\m(x) \, \bm{\Phi}(x)
      \ee
      can diverge to $+\infty$ in the limit of $g \to +\infty$.\\
      Each of these integrals can either stay finite valued or go to zero.\\
      \textbf{\textit{Insight \#3.}}\\
      Recall that $I_0=(\ref{integral-CC})$ approaches zero as $g \to +\infty$,\\
      while the integral $\int_{a}^{b} \m {\rm d}x \; \bm{\Phi}'^\dagger\n(x) \; \bm{\Phi}'\n(x)$ remains bounded (non-divergent).\\
      This means that $\bm{\Phi}(x)$ falls off to zero almost everywhere in $(a,b)$,\\
      while $\bm{\Phi}'\n(x)$ remains bounded in magnitude almost everywhere in $(a,b)$.\\
      If so, then the Wronskian (\ref{W-CC-take-2}) inevitably vanishes in the limit of $g \to +\infty$.\\
      \textbf{\textit{Insight \#4.}}\\
      If the Wronskian $W=(\ref{W-CC-take-2})\,=\,2\,i\,k_{j_{\rm in}}\,\sum_j^{E>\varepsilon_j}|T_j\n(E)|^2$
      vanishes in the limit of $g \to +\infty$,\\
      then the same must be true also for each $T_j\n(E)$ at $g \to +\infty$.\\
      Hence also the sum $\sum_j^{E>\varepsilon_j}|R_j\n(E)|^2_{g \to +\infty}=1$.\\
      {\sl Total reflection occurs in the limit of $g \to +\infty$.}\\
      \textbf{\textit{Insight \#5.}}\\
      Let $j \neq j_{\rm in}$ (here $j$ can be either an open or closed channel). We claim now that $R_j\n(E)|_{g \to +\infty}=0$.\\
      Justification: If $R_j\n(E)|_{g \to +\infty}\neq0$, then $\Psi'_j\n(x)$ would for $g \to +\infty$ exhibit a $\delta$-type jump\\
      as $x$ approaches $a$ from the right. Making the integral
      \be \label{Psi-prime-integral}
         \int_{a}^{b} \m {\rm d}x \; \bm{\Phi}'^\dagger\n(x) \; \bm{\Phi}'\n(x)
      \ee
      divergent to $+\infty$ as $g \to +\infty$. We have however excluded this possibility above in (\ref{positive-integrals-CC}).\\
      {\sl There are thus no transitions between the $j$-channels in the limit of $g \to +\infty$.}\\
      \textbf{\textit{Insight \#6.}}\\
      If $R_{j\neq j_{\rm in}}\m(E)|_{g \to +\infty}=0$, then inevitably $R_{j_{\rm in}}\m(E)|_{g \to +\infty}=e^{i\phi(E)}$\n,\\
      where $\phi(E)$ is an as yet unknown phase (possibly even $g$-dependent and $j_{\rm in}$-dependent).\\
      Thus $\Psi_{j_{\rm in}}\m(a) \, = \, e^{+ika} \, + \, e^{-i(ka-\phi(E))}$\m, and $\Psi_{j_{\rm in}}\m(a<x<b)=0$ almost everywhere.\\
      We claim now that $\phi(E)\,=\,2\,k\,a\,+\,\pi$ such that $R_{j_{\rm in}}\m(E)=-e^{+ika}$ and $\Psi_{j_{\rm in}}\m(a)=0$.\\
      The justification is analogous as in the above Insight \#5:\\
      Suppose that $R_{j_{\rm in}}\m(E)\neq-e^{+ika}$\m, then $\Psi'_{j_{\rm in}}\m(x)$ would for $g \to +\infty$ exhibit\\
      a $\delta$-type jump as $x$ approaches $a$ from the right, this would make the integral (\ref{Psi-prime-integral}) divergent.\\
      {\sl Thus, for $g \to +\infty$, the wavefunction $\Psi_{j_{\rm in}}\m(x)$ reflects at $x=a$ as a standing wave (node at $x=a$).}

\end{document}